# Lateral Heterojunction BaTiO$_3$/AlGaN Diodes with >8MV/cm Breakdown Field

Towhidur Razzak,[1,a)] Hareesh Chandrasekar,[1] Kamal Hussain,[2] Choong Hee Lee,[1] Abdullah Mamun,[2] Hao Xue,[1] Zhanbo Xia,[1] Shahadat H. Sohel,[1] Mohammad Wahidur Rahman,[1] Sanyam Bajaj,[1] Caiyu Wang,[1] Wu Lu,[1] Asif Khan,[2] Siddharth Rajan[1]

[1] Department of Electrical and Computer Engineering, The Ohio State University, Columbus, Ohio, 43210, USA

[2] Department of Electrical Engineering, University of South Carolina, 301 Main Street, Columbia SC 29208, USA

In this paper, we report enhanced breakdown characteristics of Pt/BaTiO$_3$/Al$_{0.58}$Ga$_{0.42}$N lateral heterojunction diodes compared to Pt/Al$_{0.58}$Ga$_{0.42}$N Schottky diodes. BaTiO3, an extreme dielectric constant material, has been used, in this study, as dielectric material under the anode to significantly reduce the peak electric field at the anode edge of the heterojunction diode such that the observed average breakdown field was higher than 8 MV/cm – achieved for devices with anode to cathode spacing <0.2 µm. Control Schottky anode devices (Pt/Al$_{0.58}$Ga$_{0.42}$N) fabricated on the same sample displayed an average breakdown field ~4 MV/cm for devices with similar dimensions. While both breakdown fields are significantly higher than those exhibited by incumbent technologies such as GaN-based devices, BaTiO$_3$ can enable more effective utilization of the higher breakdown fields available in ultra-wide bandgap materials by proper electric field management. This demonstration thus lays the groundwork needed to realize ultra-scaled lateral devices with significantly improved breakdown characteristics.

Significant research interest has been generated by the $Al_xGa_{1-x}N$ material system due to its attractive material properties such as extremely high critical breakdown field, predicted to be above 9 MV/cm for x ≥ 0.5, good low-field transport properties, and high electron saturation velocity comparable to GaN.[1-4] As a result, $Al_xGa_{1-x}N$ has a higher predicted Johnson Figure of Merit

$$(JFOM = \frac{v_{sat}F_{BR}}{2\pi})$$

compared to GaN. This can enable ultra-scaled devices that can surpass state of the art GaN-based amplifiers in terms of current density, breakdown voltage, and output power density at mm-wave and THz frequencies.[2-4] In addition, the large band gap of AlGaN could also enable more favorable device characteristics for normally-off devices used in power switching applications.[5]

While various figures of merit suggest the excellent potential of these materials for future device technology, approaching the breakdown limit in an actual device remains a standing challenge. Material breakdown limits for wide-band gap semiconductors have been reached previously only with PN junctions.[6] Achieving the theoretical *material* breakdown field in lateral diodes and field effect transistors is challenging due to non-uniform electric field distribution in the depletion region (which causes electric field peaking), and is further limited by tunneling breakdown at the Schottky gate electrode. These effects are more significant in the case of ultra-wide band gap semiconductors which have breakdown fields estimated to be higher than 7 MV/cm. AlGaN devices that employ metal-semiconductor based Schottky gates are limited by the breakdown strength of the Schottky diodes which typically have barrier heights of less than 2 eV. Average breakdown fields as high as 3.6 MV/cm have been demonstrated for $Al_{0.70}Ga_{0.30}N$ using metal-insulator gate electrodes, and 2.86 MV/cm for Schottky gate electrodes also for $Al_{0.70}Ga_{0.30}N$.[7,8] However, these numbers are still significantly lower than the estimated material breakdown field (11 MV/cm) at this composition. Conventional dielectrics such as $SiO_X$ and $Al_2O_3$ inserted between the gate metal and semiconductor do not offer significantly better performance since they are limited by the breakdown of the metal-dielectric junction.

More recently a solution to this problem has been proposed to be the use of extremely high permittivity oxides inserted between the metal and semiconductor layers and in the gate-drain region.[9,10] Extreme permittivity dielectrics enable improved breakdown due to two reasons. Firstly, a high dielectric constant passivation layer greatly reduces the peaking or non-uniformity in lateral electric field due to the applied drain bias and results in higher breakdown voltage and average breakdown fields due to a more uniform electric field distribution. This is illustrated in Figure 1 which shows a comparison of the electric field profiles between the anode and cathode of a heterojunction/semiconductor diode (inset Figure 1) with a (a) low-k dielectric ($\varepsilon_r$ = 10) and (b) high k dielectric ($\varepsilon_r$ = 100) for an applied anode to cathode voltage of 100 V. In addition to reducing peak fields due to depletion, the high permittivity material below the gate metal helps to reduce gate leakage which suppresses gate-leakage related breakdown. This can be understood by considering the continuity of the displacement field across the dielectric/semiconductor junction, which suppresses the electric field within the high dielectric constant material. Thus, even as voltage drops and the field increases in the semiconductor layer, the high dielectric constant layer maintains a flat potential profile, thus suppressing gate tunneling breakdown through the oxide.

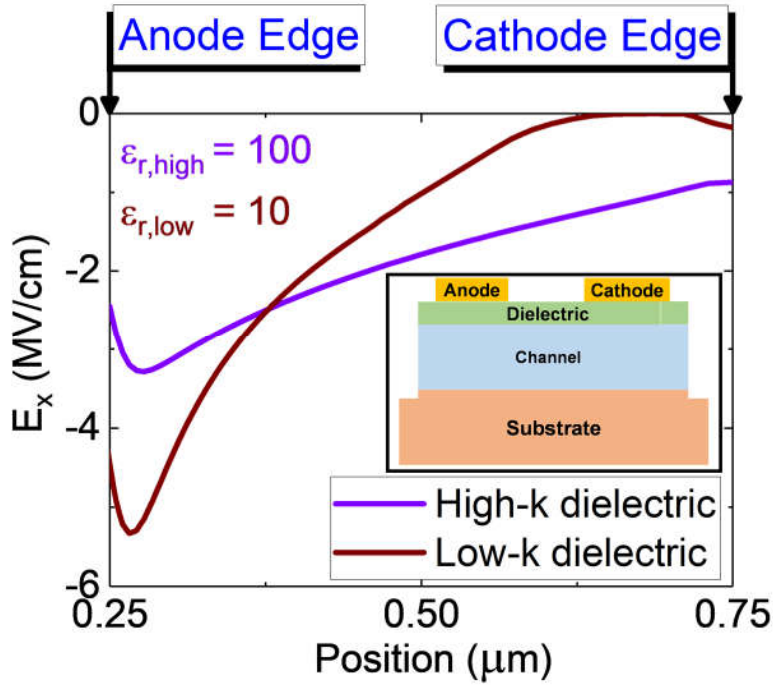

Fig. 1. Comparison of the lateral electric field profile between the anode and the cathode for heterojunction diodes with a low-k gate/anode dielectric versus a high-k gate/anode dielectric for an applied voltage, $V_{AC}$ = -100 V. Device schematic is shown on the inset.

Barium Titanate, $BaTiO_3$, a perovskite oxide displaying high dielectric constant, is a suitable candidate for such applications. In this report, we demonstrate a >8 MV/cm average breakdown field for $BaTiO_3/Al_{0.58}Ga_{0.42}N$ (predicted critical breakdown field of $Al_{0.58}Ga_{0.42}N$ ~9.7 MV/cm) heterojunction diodes which represents an important step towards achieving true material breakdown field in lateral devices. In comparison, control Schottky diodes displayed a breakdown field ~4 MV/cm, significantly lower than material breakdown field limit and representative of typical observations for Schottky devices.

Low-pressure (LP) MOCVD was used to grow the active layers for the reported device structures on 3 μm thick high quality AlN-(0001) sapphire templates. These template layers were also deposited by LPMOCVD at growth temperatures close to 1250 ℃. Their RMS surface roughness was measured by atomic-force microscope (AFM) to be 1.44 nm, as shown in Figure 2(a). The epilayer consisted of a 500 nm thick undoped i-$Al_{0.58}Ga_{0.42}N$ buffer layer grown pseudomorphically over the entire two-inch diameter AlN template at a growth temperature close to 1100 ℃. This was followed by the growth of a 60 nm thick [$Si^+$]-doped n-type $Al_{0.58}Ga_{0.42}N$ layer, with a doping concentration of $4\times10^{18}$ cm$^{-3}$ on top of the buffer layer as reported elsewhere.[11] High-resolution x-ray diffraction (XRD) spectra (BEDE D1 High Resolution Triple Axis X-ray Diffraction System) (Fig 2(b)) were used to determine the composition.

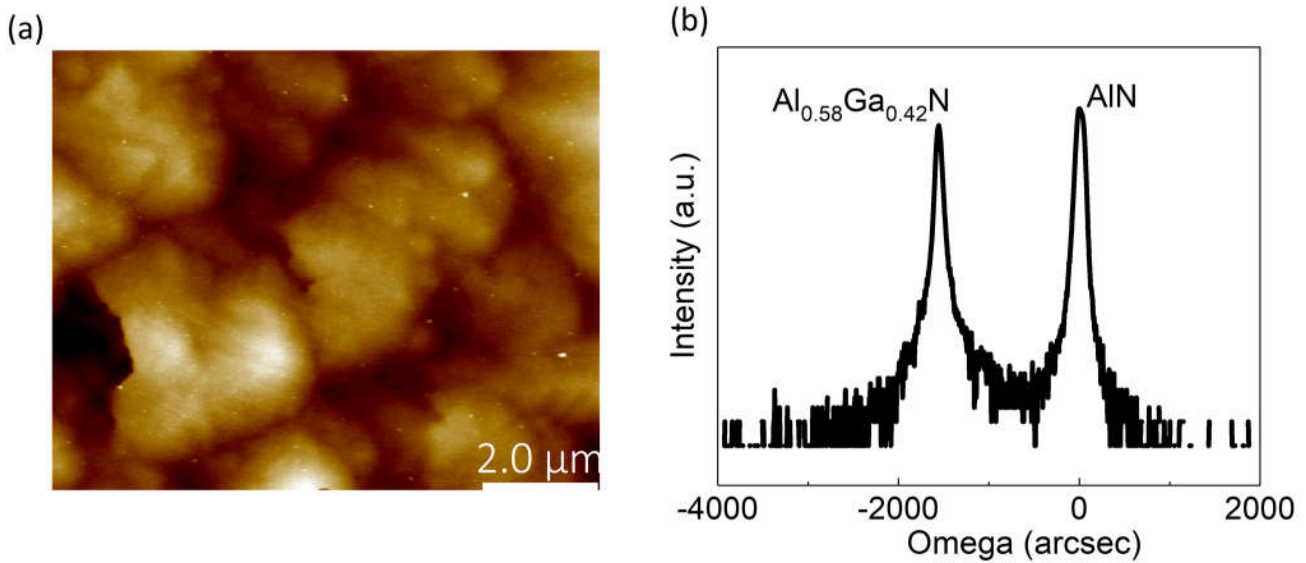

FIG. 2. (a) AFM scan of the surface of the sample showed an RMS surface roughness of 1.44 nm and (b) XRD was used to confirm the composition of the MOCVD grown film

Selective area regrowth by molecular beam epitaxy (MBE) was used to achieve ohmic contact to the MOCVD grown epilayer. 500 nm of silicon oxide ($SiO_2$) was deposited and patterned on the MOCVD-grown channel layer to act as a hard mask for MBE contact regrowth. Heavily [$Si^+$]-doped $Al_{0.58}Ga_{0.42}N$ (50 nm) was then regrown in the selectively formed pits using molecular beam epitaxy (MBE) to completely fill the etched region. This was capped with a heavily [Si+]-doped reverse Al-composition graded AlGaN layer of thickness of 50 nm and a doping of $1\times10^{20}$ cm$^{-3}$. The purpose of the reverse composition-graded n$^{++}$ AlGaN layer is to minimize abrupt conduction band offsets and facilitate low resistance non-alloyed ohmic contact formation.[8]

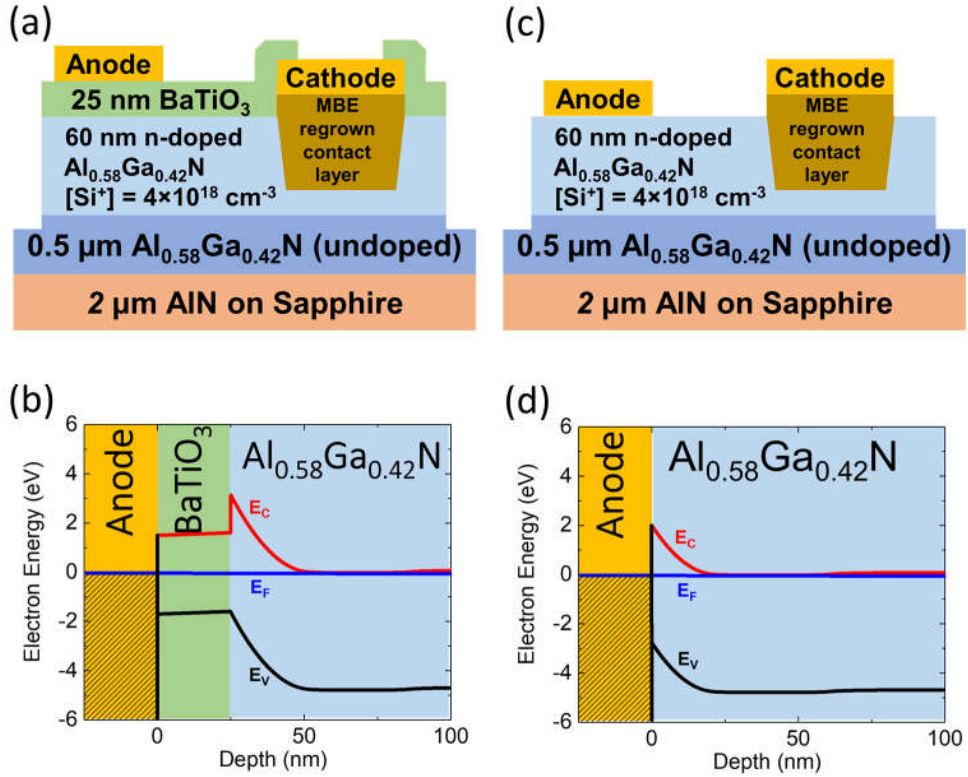

Fig. 3 (a). Schematic of the lateral BaTiO$_3$/Al$_{0.58}$Ga$_{0.42}$N heterojunction diode, (b) band-diagram under the anode for the metal/BaTiO$_3$/Al$_{0.58}$Ga$_{0.42}$N diode (c) Schematic of the fabricated lateral Al$_{0.58}$Ga$_{0.42}$N Schottky diode, and (d) Band-diagram under the anode for the Al$_{0.58}$Ga$_{0.42}$N Schottky diode

Following MBE contact regrowth, SiO$_2$ regrowth mask was removed using a diluted buffered oxide etch (BOE) solution, and a metal stack of Ti/Al/Ni/Au [20/120/30/100 nm] was deposited on the regrown contact regions via an e-beam evaporator to form ohmic contacts. This was followed by device isolation by inductively coupled plasma – reactive ion etching (ICP-RIE) plasma etching system with an RIE power of 30 W and a pressure of 5 mTorr. BaTiO$_3$ was deposited via RF sputtering at 630 ºC in an oxygen ambient. The dielectric constant of the BaTiO$_3$ films were estimated to be around 60. For the fabrication of the control samples, BaTiO$_3$ was etched away using SF$_6$ plasma ICP-RIE etching under the anode regions and the access regions while BaTiO$_3$ was etched away *only* from the access regions for the heterojunction diodes. Finally, anodes were formed by the deposition of Pt/Au [60/100 nm] metal stack via e-beam evaporation for both the Schottky and the heterojunction diodes. The schematic of the fabricated heterojunction diode is shown in Figure 3(a). Figure 3(b) shows the equilibrium energy-band diagrams calculated using self-consistent one-dimensional Poisson-Schrodinger solver under the anode for the BaTiO$_3$ sample.[12] The schematic and energy band-diagram of the control (Schottky) device are shown in Figures 3(c) and (d), respectively.

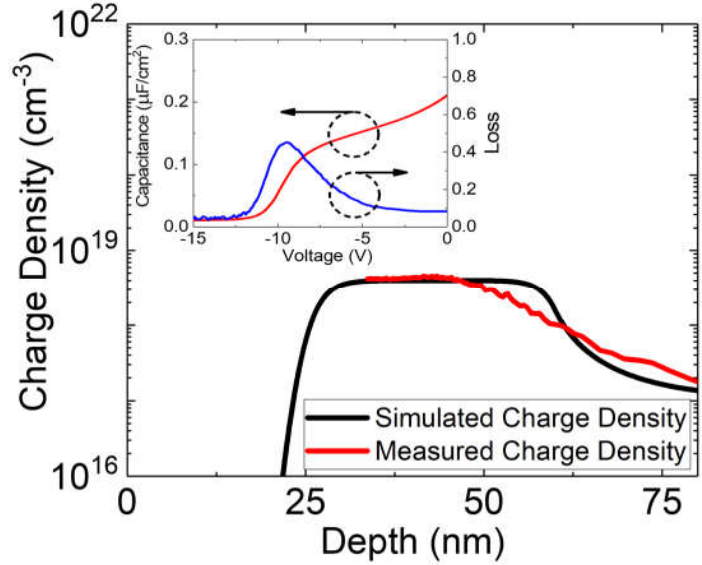

FIG. 4. Comparison of the measured charge profile to the simulated charge profile; (inset) capacitance-voltage measurement results from the BaTiO$_3$/Al$_{0.58}$Ga$_{0.42}$N diodes

Hall measurements were performed on four-terminal ungated van der Pauw (VDP) structures, and indicated sheet resistance of 6.2 kΩ/□, Hall mobility of 65 cm$^2$/V·s and a sheet carrier density of 1.55×10$^{13}$ cm$^{-2}$. Electrical characterization of the devices was performed using an Agilent B1500 parameter analyzer. Inset of Figure 4 shows the capacitance voltage characteristics measured on a 16 μm×100 μm diode. The integrated charge density, n$_s$, was found to be 1.06×10$^{13}$ cm$^{-2}$. The measured charge density and the simulated charge density are shown in Figure 4. The extracted doping density is found to be 4×10$^{18}$ cm$^{-3}$, as expected, near the top of the structure, and the lower density at larger depletion depths may be due to errors introduced by capacitor loss.

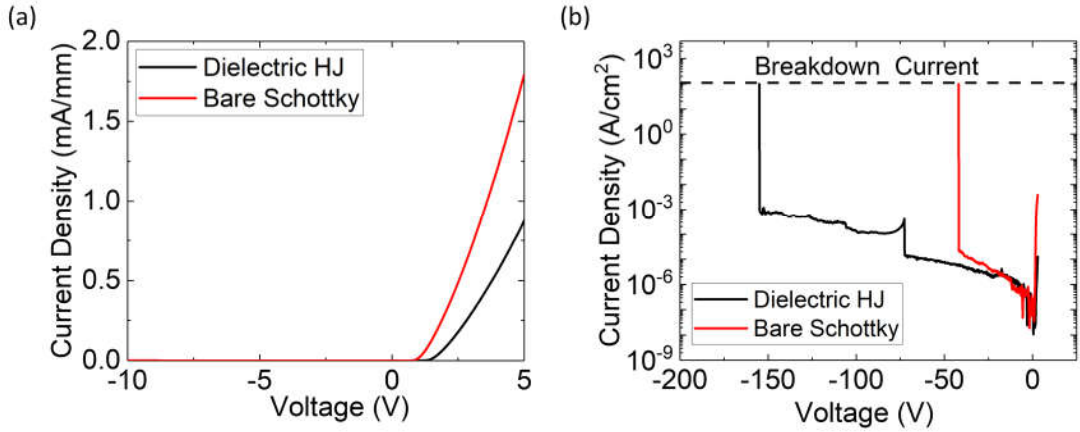

FIG. 5 Comparison between Schottky barrier diode and BaTiO$_3$/Al$_{0.58}$Ga$_{0.42}$N heterojunction diode (a) forward IV characteristics and (b) breakdown performance

Figure 5(a) shows the forward characteristics of a Schottky diode and a heterojunction diode with an anode to cathode spacing of 700 nm. The turn-on voltage of the Schottky diode and the heterojunction diode are 0.9 V and 1.5 V, respectively. Both these devices display significantly lower turn-on voltage compared to what is typically observed for AlGaN PN/PIN diodes. The differential on-resistance ($R_{ON}$), calculated from the forward IV characteristics was 31 mΩ.cm$^2$ for Schottky diode, and 56 mΩ.cm$^2$ for the heterojunction diode. The increase in $R_{ON}$ the heterojunction device is likely due to the presence of BaTiO$_3$ layer which is nominally undoped. Breakdown characteristics were evaluated for several randomly selected Schottky and heterojunction diodes. Figure 5(b) shows the breakdown characteristics for a Schottky diode and a heterojunction diode, with anode to cathode spacing of 0.19 μm and 0.18 μm respectively. While the Schottky diode breaks down at a voltage of 42 V, the heterojunction diode breaks down at a much higher voltage of 155 V despite both devices having similar anode to cathode spacing. The improvement in the breakdown field is attributed to a reduced peak field and the suppression of gate leakage current due to the BaTiO$_3$ layer. With a pinch-off voltage of -12 V, the vertical field and the horizontal field of the Schottky diode was estimated to be 3.15 MV/cm, assuming a total depleted sheet charge density of 1.55×10$^{13}$ cm$^{-2}$, and 1.58 MV/cm respectively, which gives a total field of 3.52 MV/cm. Similarly, the vertical field and the horizontal field of the heterojunction diode was estimated to 3.15 MV/cm and 7.94 MV/cm respectively, leading to an effective electric field of 8.5 MV/cm. Multiple randomly chosen devices were measured for both the Schottky diodes and the heterojunction diodes and the results are show in Figure 6(a). In general, for the heterojunction diodes, for an anode to cathode spacing in the range of 0.2-0.3 μm, the minimum breakdown fields observed were in the range of 6-8 MV/cm. Figure 6(b) shows the highest reported breakdown fields for various semiconductor materials as a function of the bandgap. The value reported in this work, 8.5 MV/cm, is the highest reported experimental breakdown field for any semiconductor material.

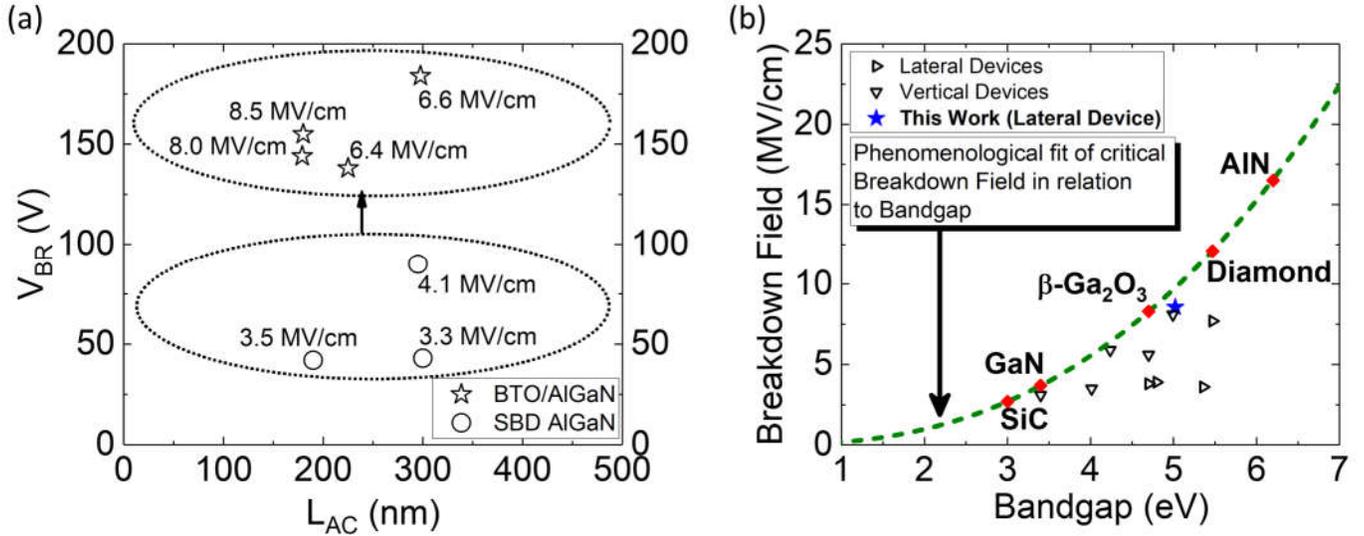

FIG. 6. (a) Breakdown performance for lateral $BaTiO_3/Al_{0.58}Ga_{0.42}N$ heterojunction diodes and AlGaN Schottky barrier devices for $L_{AC}$ < 500 nm and (b) comparison of breakdown fields in various semiconductor devices achieved to date [6,7,9,13-17]. The dashed green line shows the phenomenological fit of critical breakdown field as a function of bandgap. The red dots represent the predicted maximum breakdown field achievable for various materials.

The breakdown voltage achieved for the diode with average breakdown field of 8.5 MV/cm was 155 V, which is lower than the theoretically estimated maximum breakdown field of 9.7 MV/cm and a flat electric field distribution. Therefore, there exists room for optimization of this $BaTiO_3$/AlGaN heterostructure design including optimization of $BaTiO_3$ thickness and growth conditions together with the exploration of potential integration of field plate structures. One important benefit of using $BaTiO_3$ as a gate or anode dielectric is that it enables the fabrication of ultra-scaled devices while meeting the breakdown voltage requirements due to the much higher breakdown fields achievable in these structures. This is particularly beneficial for low mobility materials like AlGaN and $Ga_2O_3$ which requires scaled devices for high performance RF electronics to compensate for the higher sheet resistance compared to GaN. The presence of the high-k dielectric also leads to scaling of the gate to channel distance which can result in lower output conductance and higher transconductance.[10] However, it should be noted that in such a heterojunction with a high-k dielectric, the drain depletion region is longer than the case with a low-k dielectric and correspondingly higher gate to channel capacitance due to the high-permittivity region between the gate and drain. As a result, the peak cutoff frequency is expected to be lower than a device with low-k barrier.[10] It is therefore important to properly optimize the thickness of the high-permittivity region to manage the tradeoff between field management and gate–drain capacitance dielectric. Scaling, enabled by the high-k heterostructure devices, is also expected to play an important role for next generation power electronics, whereby the on-resistance of the device can be significantly reduced, due to shortened device dimensions while at the same time delivering higher output power more efficiently.

In conclusion, we have demonstrated Pt/BaTiO$_3$/Al$_{0.58}$Ga$_{0.42}$N lateral heterojunction diodes with enhanced breakdown characteristics. By using BaTiO$_3$ as a gate and access-region dielectric, the peak field at the anode edge was significantly reduced and an average breakdown field exceeding 8 MV/cm was achieved for devices with anode to cathode spacing of 0.18 μm. In contrast, Pt/Al$_{0.58}$Ga$_{0.42}$N control Schottky diodes displayed an average breakdown field below 4 MV/cm for devices with similar dimensions. The use of a high-k dielectric can more effectively utilize the high breakdown fields in ultra-wide bandgap materials, by proper electric field management. This demonstration thus provides a framework to realize ultra-scaled lateral devices with improved breakdown characteristics.

**ACKNOWLEDGMENTS**

The authors acknowledge funding from Air Force Office of Scientific Research (AFOSR Grant FA9550-17-1-0227, Program Manager Kenneth Goretta) and the DARPA DREaM program (ONR N00014-18-1-2033, Program Manager Dr. Young-Kai Chen, monitored by Office of Naval Research, Program Manager Dr. Paul Maki).